\documentclass[pra,aps,showpacs,reprint]{revtex4-1}
\usepackage{}
\usepackage{mathrsfs}
\usepackage{amsfonts}
\usepackage{txfonts}
\usepackage{amssymb}
\usepackage{graphicx}
\usepackage{bm}
\newcommand{\ket}[1]{|#1\rangle}
\newcommand{\bra}[1]{\langle #1|}

\begin{document}
\title{Non-adiabatic Holonomic Gates realized by a single-shot implementation}
\author{G. F. Xu}
\affiliation{Department of Physics, Shandong University, Jinan
250100, China}
\author{C. L. Liu}
\affiliation{Department of Physics, Shandong University, Jinan
250100, China}
\author{P. Z. Zhao}
\affiliation{Department of Physics, Shandong University, Jinan
250100, China}
\author{D. M. Tong}
\email{tdm@sdu.edu.cn}
\affiliation{Department of Physics, Shandong University, Jinan
250100, China}
\date{\today}
\begin{abstract}
Non-adiabatic holonomic quantum computation has received increasing attention due to its robustness against control errors. However, all the previous schemes have to use at least two sequentially implemented gates to realize a general one-qubit gate. In this paper, we put forward a novelty scheme by which one can directly realize an arbitrary holonomic one-qubit gate with a single-shot implementation, avoiding the extra work of combining two gates into one. Based on a three-level model driven by laser pulses, we show that any single-qubit holonomic gate can be realized  by varying the detuning, amplitude, and phase of lasers. Our scheme is compatible with previously proposed non-adiabatic holonomic two-qubit gates, combining with which the arbitrary holonomic one-qubit gates can play universal non-adiabatic holonomic quantum computation. We also investigate the effects of some unavoidable realistic errors on our scheme.

\end{abstract}
\maketitle
\date{\today}

\section{Introduction}
Quantum computation uses quantum logic totally different from the
Boolean logic on which classical computation are built. By using the
quantum parallelism, quantum computation is believed to be qualitatively faster than classical
computation in solving many problems.
Circuit-based quantum computation relies on the ability to perform a universal high-fidelity quantum gates. Two main challenges in achieving such high-fidelity gates are to reduce control errors of a quantum system and to avoid decoherence between the system and its environment. To overcome the problems, various proposal of fault-tolerant quantum computation are proposed. A promising one of such proposals is non-adiabatic holonomic quantum computation, being with fault-tolerant feature.

Non-adiabatic holonomic quantum computation is based on non-adiabatic and non-Abelian geometric phases. It may be taken as an extension of traditional geometric quantum computation based on adiabatic and Abelian geometric phases.
Early in 1984, Berry found that a quantum system in a non-degenerate eigenstate undergoing adiabatic cyclic
evolution acquires a geometric phase (adiabatic and Abelian geometric phase) \cite{Berry}. Soon after, the notion of geometric
phase originally for quantum systems with non-degenerate eigenstates in adiabatic evolutions
was gradually generalized to quantum systems with degenerate eigenstates in adiabatic evolutions (adiabatic and non-Abelian geometric phase, i.e. adiabatic quantum holonomy) \cite{Wilczek} , quantum systems with non-degenerate eigenstates in non-adiabatic evolutions (non-adiabatic and Abelian geometric phase) \cite{Aharonov}, and quantum systems with degenerate eigenstates in non-adiabatic evolutions (non-adiabatic and non-Abelian geometric phase, i.e. non-Adiabatic holonomy) \cite{Anandan}, respectively. Since geometric phases, both Abelian and non-Abelian, are only dependent on the path (subspace) in which the system evolves but independent of its evolutional details, quantum computations based on geometric phases are robust against certain control errors. The first scheme of geometric quantum computation is based on adiabatic and Abelian geometric phases, proposed by Jones et al \cite{Jones}. It immediately prompted quantum computation based on adiabatic and non-Abelian geometric phase, i.e. adiabatic holonomic quantum computation \cite{zanardi99,Duan}, and quantum computation based on non-adiabatic and Abelian geometric phase \cite{WangXB,Zhu1}.
Recently, the proposal of quantum computation based on non-adiabatic and non-Abelian geometric phase, i.e. non-adiabatic holonomic quantum computation, was found too \cite{Sjoqvist,Xu}.

Non-adiabatic holonomic quantum computation shares all the geometric nature of its adiabatic counterpart while avoids the long run-time requirement. Although the proposal of non-adiabatic holonomic quantum computation was only recently proposed, it has received increasing attention due to its robustness  against control errors and its rapidity without the speed limit of  the adiabatic evolution. A number of alternative theoretical and experimental schemes are prompted   \cite{Johansson,Abdumalikov,Feng,Spiegelberg,Mousolou,Mousolou1,Zhang1,Arroyo,Zu,Xu1,Xu2,Liang,Zhang2,Zhou,Xue,Sjoqvist1,Pyshkin,Song}.  Specially, non-adiabatic holonomic quantum computation has been experimentally demonstrated with a three-level transmon qubit \cite{Abdumalikov}, with a NMR quantum information processor \cite{Feng}, and with diamond nitrogen-vacancy centers \cite{Arroyo,Zu}, successively. However, all the previous schemes of non-adiabatic holonomic quantum computation have to use at least two sequentially implemented gates to realize a general one-qubit gate. In this paper, we put forward a novelty scheme by which one can directly realize an arbitrary holonomic one-qubit gate with a single-shot implementation, avoiding the extra work of combining two gates into one. We also investigate the effects of some unavoidable realistic errors on the gates. Our scheme is compatible with the previously proposed two-qubit non-adiabatic holonomic gate, and therefore our arbitrary one-qubit holonomic gates combining with the previous two-qubit gate can realize universal non-adiabatic holonomic quantum computation.

The structure of this paper is organized  as follows. In Sec. II, we elucidate the physical model and the method to realize an arbitrary holonomic  one-qubit quantum gate with a single-shot implementation. In Sec. III, we investigate the effects of some unavoidable realistic errors on the gates. Section IV is the discussion and conclusion.

\section{The single-shot scheme}
\begin{figure}[htbp]
 \begin{center}
\includegraphics[width=5.5cm, height=4.6cm]{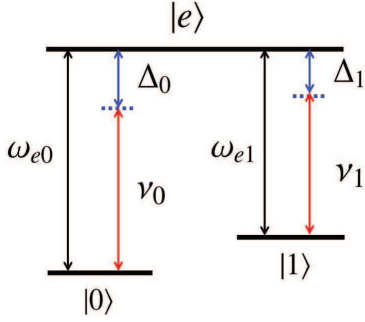}
 \end{center}
\caption{(color online). The level configuration with
driving laser fields. $\omega_{ej}=\omega_e-\omega_j$ is the energy difference between states $\ket{e}$ and $\ket{j}$.
$\nu_j$ is the oscillation frequency of laser field driving the transition $\ket{j}\leftrightarrow\ket{e}$
and $\Delta_j=\nu_j-\omega_{ej}$ is the corresponding detuning.}\label{fig1}
\end{figure}

Let us first elucidate the physical model. Consider a three-level quantum system with driving laser fields. The eigenvalues and corresponding eigenstates of the system are denoted as $\omega_0$, $\omega_1$, $\omega_e$, and $\ket{0}$, $\ket{1}$, $\ket{e}$, respectively.
The laser field driving the transition $\ket{j}\leftrightarrow\ket{e}$ is described by ${\bf E}_j(t)={\bm \epsilon}_jg_j(t)\cos\nu_jt$,
where ${\bm \epsilon}_j$ is the polarization, $g_j(t)$ is the envelope function, and $\nu_j$ is the oscillation frequency. In our scheme, the logic space is spanned by $\ket{0}$ and $\ket{1}$,  and $\ket{e}$ as an ancillary state. Figure \ref{fig1} illustrates the level configuration of the system with the driving laser fields.
Hamiltonian of the three-level system with the driving laser fields  reads,
with setting the energy of state $\ket{e}$ to zero,
\begin{eqnarray}
\mathcal {H}(t)=H_0+{\bm \mu}\cdot[{\bf E}_0(t)+{\bf E}_1(t)], \label{hamiltonianlab}
\end{eqnarray}
where $H_0=-\omega_{e0}\ket{0}\bra{0}-\omega_{e1}\ket{1}\bra{1}$ with $\omega_{ej}=\omega_e-\omega_j$ is the bare Hamiltonian, and ${\bm \mu}$ is the electric dipole operator. The three-level model is very common in many physical systems, and it is has been widely used in holonomic quantum computaion \cite{Sjoqvist,Johansson,Abdumalikov,Spiegelberg,Arroyo,Zu}.

Based on the above three-level model, we will show that any holonomic one-qubit gate can be realized by separately varying the detuning, amplitude, and phase of lasers. To this end, we address the issue step by step.

First, we simplify the Hamiltonian sequentially by taking the rotating wave approximation and by properly setting the parameters in it

By using the rotating frame with transform operator $V(t)=\exp \left[-i\left(\nu_0\ket{0}\bra{0}+\nu_1\ket{1}\bra{1}\right)t\right]$, and ignoring rapidly oscillating terms in the rotating wave approximation, the Hamiltonian can be written as
\begin{eqnarray}
\mathcal {H}_{rot}(t)=\sum_{j=0}^1\Delta_j\ket{j}\bra{j}+\big(\Omega_j(t)\ket{j}\bra{e}+h.c.\big), \label{hamiltonian}
\end{eqnarray}
with
\begin{eqnarray}
\Delta_j=\nu_j-\omega_{ej},~~~\Omega_j(t)=g_j(t)\bra{e}{\bm \mu\cdot{\bm \epsilon}_j}\ket{j}, \label{set1}
\end{eqnarray}
 where $\Delta_j$ and $\Omega_j(t)$ are the detuning and the pulse envelope respectively, and $h.c.$ means hermitian conjugate. Note that $\Omega_j(t)\ll\nu_j$ is necessary for the validity of this approximation.

By setting the the detunings to
\begin{eqnarray}
\Delta_0=\Delta_1=\Delta,
\label{deltadelta}
\end{eqnarray}
$\mathcal {H}_{rot}(t)$ can be simplified as
\begin{eqnarray}
\mathcal {H}_{rot}(t)=\Delta\big({I}-\ket{e}\bra{e}\big)+\sum_{j=0}^1\big(\Omega_j(t)\ket{j}\bra{e}+h.c.\big),
\end{eqnarray}
where $I=\ket{e}\bra{e}+\ket{0}\bra{0}+\ket{1}\bra{1}$, being an identity operator for the three-level system. Note that the term $\Delta{I}$ in the Hamiltonian generates only a global phase during the evolution, which does not affect the quantum gates. Hence, the gate is only determined by the effective  Hamiltonian,
\begin{eqnarray}
\mathcal {H}_{eff}(t)=-\Delta\ket{e}\bra{e}+\sum_{j=0}^1\big(\Omega_j(t)\ket{j}\bra{e}+h.c.\big).
\label{Heff}
\end{eqnarray}

We suppose that the transition $\ket{j}\leftrightarrow\ket{e}$ is driven by the square pulse, and set it as
\begin{eqnarray}
\left\{\begin{array}{c}
\Delta=-2\Omega\sin\gamma, \\
\Omega_0(t)=\Omega\cos\alpha\cos\gamma,\\
\Omega_1(t)=\Omega{e^{i\beta}}\sin\alpha\cos\gamma,
\end{array}\right.\label{set2}
\end{eqnarray}
where $\Omega$ can be regarded as the norm of the vector $(\Delta/2, \Omega_0(t), \Omega_1(t))$, such that $\mathcal {H}_{eff}(t)$ becomes
\begin{eqnarray}
\mathcal {H}_{eff}&=&\Omega\sin\gamma(\ket{e}\bra{e}+\ket{b}\bra{b})+\Omega[\cos\gamma(\ket{b}\bra{e} \nonumber \\
&&+\ket{e}\bra{b})+\sin\gamma(\ket{e}\bra{e}-\ket{b}\bra{b})], \label{hamiltoniang}
\end{eqnarray}
where $\ket{b}=\cos\alpha\ket{0}+e^{i\beta}\sin\alpha\ket{1}$.  It is worth to note that $\Delta$ and $\Omega_j(t)$ are respectively determined by the frequencies and amplitudes of the lasers. It is difficult to make them have the same envelope except for using the square pulses. This is the reason that we choose the square pulses here. The state orthogonal to $\ket{b}$ will be denoted as  $\ket{d}=\sin\alpha\ket{0}-e^{i\beta}\cos\alpha\ket{1}$. $\{\ket{b}, \ket{d}\}$ spans the same subspace as that by $\{\ket{0}, \ket{1}\}$. Since $\Omega_j(t)\ll\nu_j$ in the rotating wave approximation, it is reasonable to further assume the detuned
driving laser fields do not induce other unexpected transitions of the three-level system except those described in Eq. (\ref{hamiltoniang}).
Hamiltonian $\mathcal {H}_{eff}$ in Eq. (\ref{hamiltoniang}) is realizable.

Second, we construct qubit gates by the aid of the effective Hamiltonian $\mathcal {H}_{eff}$.

Equation (\ref{hamiltoniang}) shows that only states $\ket{e}$ and $\ket{b}$ are coupled by the dynamics, while the third state $\ket{d}=\sin\alpha\ket{0}-e^{i\beta}\cos\alpha\ket{1}$ decouples from the dynamics. In this case, we can express $\mathcal {H}_{eff}$ in the basis of the spin-half operators. With mapping $\ket{e}\bra{e}+\ket{b}\bra{b}\longrightarrow{I_2}$, $\ket{e}\bra{b}+\ket{b}\bra{e}\longrightarrow\sigma_x$, and $\ket{e}\bra{e}-\ket{b}\bra{b}\longrightarrow\sigma_z$, the effective Hamiltonian can be simply expressed as
\begin{eqnarray}
\mathcal {H}_{eff}=\Omega\sin\gamma I_2+\Omega(\cos\gamma\sigma_x+\sin\gamma\sigma_z). \label{hsigma}
\end{eqnarray}
With the above expression of $\mathcal {H}_{eff}$, it is easy to calculate $U(t)=e^{-iH_{eff}t}$ . If the evolution period $T$ is taken as
\begin{eqnarray}
T=\frac{\pi}{\Omega},\label{time}
\end{eqnarray}
then the resulting evolution operator, in the basis $\{\ket{e}, \ket{b}, \ket{d}\}$, can be written as
\begin{eqnarray}
U(T)= \left(\begin{array}{ccc}
e^{-i\phi} & 0 & 0 \\
0 & e^{-i\phi} & 0 \\
0 & 0 & 1
\end{array}\right), \label{ugt}
\end{eqnarray}
where $\phi=\pi\sin\gamma+\pi$.
When we restrict the attention only to the logical subspace spanned by $\{\ket{0}, \ket{1}\}$,  the
evolution operator $U(T)$ is equivalent to
\begin{eqnarray}
U_L(T)=e^{-i\frac{\phi}{2}(\ket{b}\bra{b}-\ket{d}\bra{d})}.
\label{unitary}
\end{eqnarray}
In the above, the rotation angle of the qubit gate is determined by $\sin\gamma$, while the rotation axis of it is determined by $\ket{b}$ and $\ket{d}$.
To explain how one can obtain any qubit gate by varying the detuning, amplitude, and phase of lasers, we recall the relations in Eqs. (\ref{set1}), (\ref{deltadelta}), (\ref{set2}) and (\ref{time}). If we want to obtain a gate $U_L(T)=e^{-i\frac{\phi}{2}(\ket{b}\bra{b}-\ket{d}\bra{d})}$, we may first calculate $\alpha$, $\beta$ and $\gamma$, and then determine $\Omega_0(t)$, $\Omega_1(t)$ and $\Delta$ by using Eqs. (\ref{set2}) and (\ref{time}). With the known $\Omega_0(t)$, $\Omega_1(t)$ and $\Delta$, the laser parameters $g_j(t)$, $\nu_j$ and the detuning $\Delta_j$ can be properly chosen by the aid of Eqs. (\ref{set1}) and (\ref{deltadelta}).

Third, we demonstrate that the quantum gate defined in Eq. (\ref{unitary}) is a holonomic matrix in the subspace spanned by $\ket{0}$ and $\ket{1}$.

Before proceeding further, it is instructive to recapitulate what evolution operator of an $N$-dimensional quantum system plays a holonomy transformation in an $L$-dimensional subspace. For an $N-$dimensional quantum system defined by $H(t)$, if there is a time-dependent $L-$dimensional subspace $\mathcal{S}(t)$ spanned by the orthonormal vectors $\{ \ket{\phi_k(t)} \}_{k=1}^L$ that satisfy $i\ket{\dot\phi_k(t)}=H(t)\ket{\phi_k(t)}$, i.e. $\ket{\phi_k(t)} = U_H(t)\ket{\phi_k(0)}$ with $U_H(t) =
{\bf T} \exp{i\int_0^tH(t')dt'}$, and if $\ket{\phi_k (t)}$ satisfy the two
conditions: $ \textrm{(i)}$  $\sum_{k=1}^L |\phi_{k}(T)\rangle \langle
\phi_{k}(T)| =
\sum_{k=1}^L |\phi_{k}(0)\rangle \langle \phi_{k}(0)|$,  and $\textrm{(ii)}$  $\langle
\phi_{k}(t)|H(t)|\phi_{l}(t)\rangle=0,\ k,l=1,...,L,$
then the unitary transformation $U_H(T)$ is a holonomy matrix on the $L-$dimensional subspace $\mathcal{S}(0)$ spanned by $\{ \ket{\phi_k(0)} \}_{k=1}^L$.
Condition $\textrm{(i)}$ indicates that the evolution of the subspace $\mathcal{S} (t)$ is cyclic, while condition $\textrm{(ii)}$ guarantees that
there is no dynamical component in the cyclic evolution.

We now show that for the system governed by $\mathcal {H}_{eff}$, the evolution operator $U(T)$ defined in Eq. (\ref{ugt}) plays a holonomy gate in the logic subspace.
To this end, we examine the validity of the conditions $\textrm{(i)}$ and $\textrm{(ii)}$  for this system. It is clear that $U(T)(\ket{0}\bra{0}+\ket{1}\bra{1})U^{\dag}(T)=\ket{0}\bra{0}+\ket{1}\bra{1}$, i.e. the first condition is fulfilled.
Since $\mathcal {H}_{eff}$ is time-independent and the subspace spanned by $\{\ket{b}, \ket{d}\}$ is the same as that by $\{\ket{0}, \ket{1}\}$, condition $\textrm{(ii)}$ reduces to
$\bra{m}\mathcal {H}_g(t)\ket{n}=0$, where $m,n\in\{b, d\}$.  The validity of the second condition is easily verified by immediately substituting Eq. (\ref{hamiltoniang}) into the condition.
 Therefore, the evolution operator $U(T)$ plays a holonomy gate in the logic subspace, for the system governed by the effective Hamiltonian $\mathcal {H}_{eff}$.

\section{Effects of realistic errors}

Holonomic quantum gates are robust against the control errors that change the evolutional rates of systems but keep the evolutional paths unchanged. However, they may be still sensitive to those errors that affect the evolutional paths. In this section, we investigate the effects of some unavoidable realistic errors on our scheme. As known, there are many different sources of imperfections, and the impacts of them are specific to the particular system that is used for the implementation of the gates. Here, we therefore restrict our attention to some general sources of errors
that are typically encountered in a variety of implementations.
Specifically, for the errors induced by the environment, we consider dephasing, and for the errors induced by the external driving fields, we consider pulse area error and frequency detuning error. To quantify the performance of the gates, we use the average fidelity $\overline{F}$, which equals the average value of
the fidelity $F=Tr(\rho_{out}\rho^\prime_{out})$ for an input state,
where $\rho_{out}$ and $\rho^\prime_{out}$ are respectively the ideal and real output states for the input state.
Without loss of generality, the input state can be expressed as
$\ket{\varphi_{in}}=\cos\frac{\theta}{2}e^{i\frac{\varphi}{2}}\ket{b}
+\sin\frac{\theta}{2}e^{-i\frac{\varphi}{2}}\ket{d}$,
where $\theta\in[0,\pi]$ and $\varphi\in[0,2\pi]$ are the polar and
azimuthal angles respectively.

Dephasing is caused by the inevitable interaction of the system with its environment. For a variety of systems, dephasing is the major source of decoherence. We suppose that the Markovian approximation is valid for the system and the effect of
dephasing can be described by the Lindblad equation
\begin{eqnarray}
\frac{d\rho}{dt}=-i[\mathcal {H}_{eff},\rho]
+\sum_{k=0,1}(2L_{k}\rho{L}_{k}^\dag-L_{k}^\dag{L_{k}}\rho-\rho{L}_{k}^\dag{L_{k}}),
\end{eqnarray}
where $\rho$ is the density matrix of the system, and $L_k=\sqrt{\epsilon}(\ket{e}\bra{e}-\ket{k}\bra{k})$ is the Lindblad operator with $\epsilon$ being the coupling parameter.
\begin{figure}[htbp]
 \begin{center}
\includegraphics[width=8.0cm, height=5.0cm]{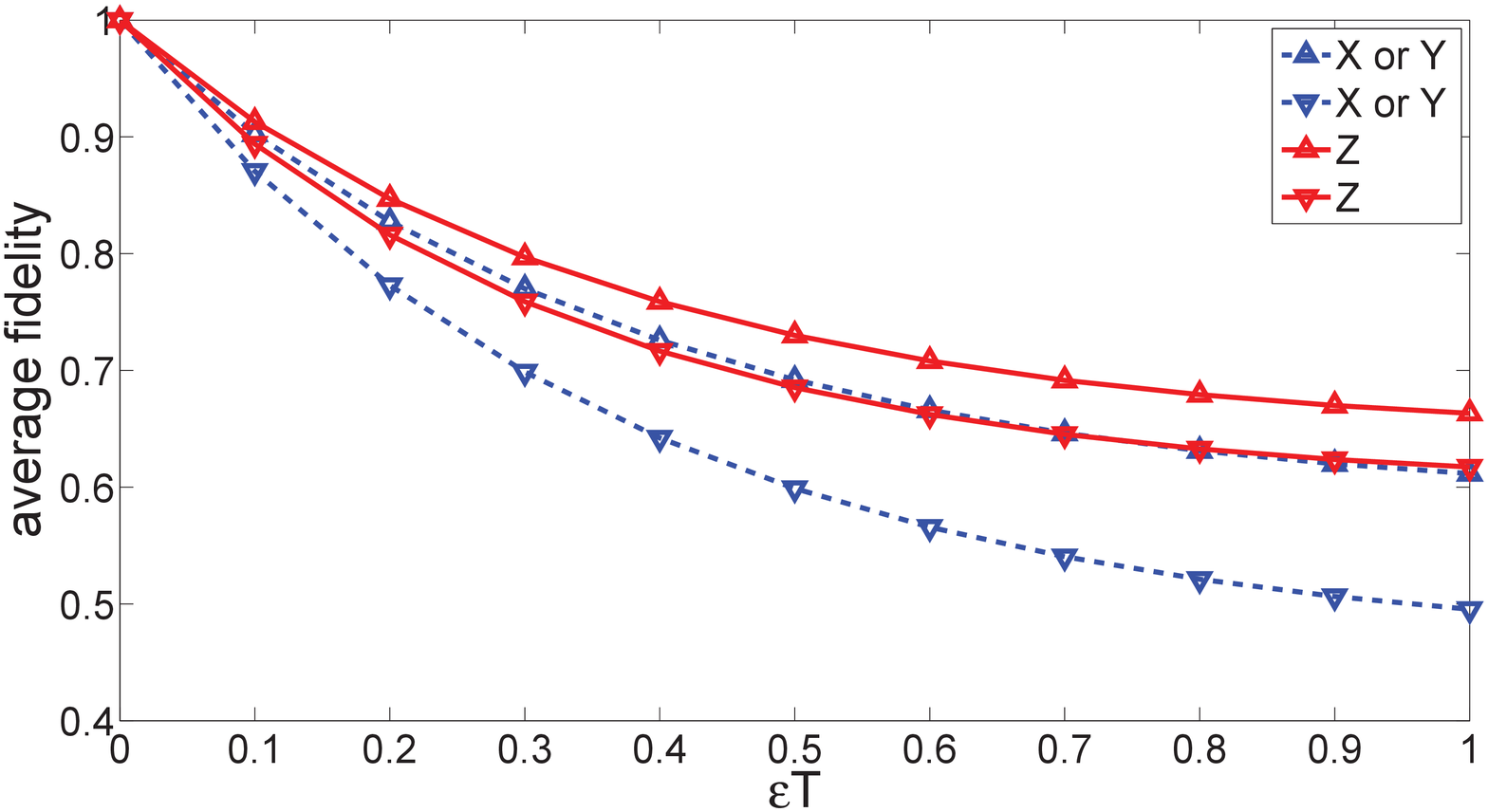}
 \end{center}
\caption{(color online). Effects of dephasing on rotation gates about $X$, $Y$ and $Z$ axes. Blue dashed lines and red solid
lines respectively represent gates about $X$ (or $Y$) and $Z$ axes. Upper triangle and lower triangle respectively represent rotation angles ${\pi}/{6}$ and ${\pi}/{3}$.}\label{fig2}
\end{figure}
Figure \ref{fig2} is our numerical result, which illustrates the performance of the rotation gates about $X$, $Y$ and $Z$ axes. It shows that the average fidelities decrease with the increasing of the error parameter $\varepsilon{T}$ and rotation angle. The rotation gate about $Z$ axis is more robust than that about $X$ or $Y$ axis. Specially, the average fidelities of rotation gates about $X$ and $Y$ axes are the same. In fact, the average fidelities of all the rotation gates about axes with the same parameter $\alpha$ (the parameter $\beta$ can be different) are equal to each other since their evolutions can be connected by the unitary operators $\exp[i{\varphi}(\ket{0}\bra{0}-\ket{1}\bra{1})/2]$.

The pulse area error is usually caused by inaccurate control in the laser pulse duration. We assume the two driving laser fields have the same type of pulse area error, and the error Hamiltonian describing the three-level system
can be expressed as
\begin{eqnarray}
\mathcal {H}_p(t)=\mathcal {H}_{eff}(t)+\mathcal {H}_{\xi}(t),
\end{eqnarray}
where $\mathcal {H}_{eff}(t)$ is the effective Hamiltonian described by Eq. (\ref{Heff}), and $\mathcal {H}_{\xi}(t)=\xi\sin\gamma(\ket{e}\bra{e}+\ket{b}\bra{b})+\xi[\cos\gamma(\ket{b}\bra{e}
+\ket{e}\bra{b})+\sin\gamma(\ket{e}\bra{e}-\ket{b}\bra{b})]$ is the error Hamiltonian, with $\xi$ being the error parameter.
For the above Hamiltonian, the fidelity reads
\begin{eqnarray}
F_p&=&\big[\sin^2\frac{\theta}{2}+\cos^2\frac{\theta}{2}(\cos\xi^\prime{T}\cos\xi{T}+\sin\xi^\prime{T}\sin\xi{T}\sin\gamma)\big]^2 \nonumber \\
&&+\cos^4\frac{\theta}{2}\big[\cos\xi^\prime{T}\sin\xi{T}\sin\gamma-\sin\xi^\prime{T}\cos\xi{T}\big]^2,
\end{eqnarray}
where $\xi^\prime=\xi\sin\gamma$.
The above expression shows that the fidelity $F_p$ is independent of the input state parameter $\varphi$. In general,  $\xi{T}$, describing the error, is expected to be small, and the fidelity $F_p$ can be approximately expressed as
\begin{eqnarray}
F_p\approx{1-\cos^2\frac{\theta}{2}\cos^2\gamma(\xi{T})^2}. \label{f1}
\end{eqnarray}
The average fidelity takes the following expression,
\begin{eqnarray}
\overline{F}_p\approx1-\frac{1}{2}\cos^2\gamma(\xi{T})^2. \label{f1ave}
\end{eqnarray}
Equations (\ref{f1}) and (\ref{f1ave}) show that $F_p$ and $\overline{F}_p$ decrease with the increasing of the error parameter $|\xi{T}|$ but increase with the increasing of the rotation angle parameter $|\sin\gamma|$. For a particular gate, while the pulse area error
does not affect state $\ket{d}$, it has the most detrimental effect on state $\ket{b}$. However, the average fidelity $\overline{F}_p$ is independent of the rotation axes of the gates.

The frequency detuning error is usually caused by the frequency difference between the ideal and real driving
laser fields. For simplicity, we assume that the two driving laser fields have the same detuning error. The new detuning gives rise to additional diagonal term in the Hamiltonian, and the error Hamiltonian can be written as
\begin{eqnarray}
\mathcal {H}_f(t)=\mathcal {H}_{eff}(t)+\kappa(\ket{0}\bra{0}+\ket{1}\bra{1}),
\end{eqnarray}
where $\kappa$ is the error parameter. Note that the above Hamiltonian is in the frame co-rotating with the new detuned laser fields.
With detailed calculations, one can get the expression of the fidelity,
\begin{eqnarray}
F_f&=&\big(\sin^2\frac{\theta}{2}-\cos^2\frac{\theta}{2}\cos\frac{\kappa{T}}{2}\cos{BT} \nonumber \\
&&+\cos^2\frac{\theta}{2}\sin\frac{\kappa{T}}{2}\sin{BT}\frac{\Omega\sin\gamma-\frac{\kappa}{2}}{B}\big)^2 \nonumber \\
&&+\big(\cos^2\frac{\theta}{2}\sin\frac{\kappa{T}}{2}\cos{BT} \nonumber \\
&&+\cos^2\frac{\theta}{2}\cos\frac{\kappa{T}}{2}\sin{BT}\frac{\Omega\sin\gamma-\frac{\kappa}{2}}{B}\big)^2,
\end{eqnarray}
where $B=\sqrt{\Omega^2-\kappa\Omega\sin\gamma+\frac{\kappa^2}{4}}$.
Again, the fidelity is independent of the input state parameter $\varphi$. In general, $\kappa{T}$, describing the detuning error, is small, and then the fidelity can be approximately expressed as
\begin{eqnarray}
F_f\approx1-\frac{1}{4}\cos^2\frac{\theta}{2}\cos^2\gamma\big(1-\cos^2\frac{\theta}{2}\cos^2\gamma\big)(\kappa{T})^2. \label{f2}
\end{eqnarray}
The average fidelity can be then obtained as
\begin{eqnarray}
\overline{F}_f\approx1-\frac{1}{24}\cos^2\gamma(3-2\cos^2\gamma)(\kappa{T})^2. \label{f2ave}
\end{eqnarray}
 Equations (\ref{f2}) and (\ref{f2ave}) show that both $F_f$ and $\overline{F}_f$ decrease with the increasing of the error parameter $|\kappa{T}|$. While the average fidelity $\overline{F}_f$ has the minimum value when the rotation angle parameter $|\sin\gamma|=\frac{1}{2}$,
 the relationship between the fidelity $F_f$ and the rotation angle parameter is not deterministic, which depends on the value $\frac{1}{1+\cos\theta}$.  The most robust state is still state $\ket{d}$, similar to the case of the pulse area error. However, the most sensitive state is not $\ket{b}$ but the state satisfying the condition $\cos^2\frac{\theta}{2}=\frac{1}{2\cos^2\gamma}$. The average fidelity $\overline{F}_f$ is still independent of the rotation axes of the gates.

\section{Discussion and conclusion}

It is worth of noting that our scheme is compatible with the previously proposed non-adiabatic holonomic two-qubit gates. To realize universal quantum computation, the arbitrary one-qubit gates in our scheme need to combine with a nontrivial two-qubit gate. Note that a number of universal quantum computation schemes with nontrivial non-adiabatic holonomic two-qubit gates have been proposed in the previous works, for instances, the well-known scheme in Ref. \cite{Sjoqvist} and the experimental implementation in Ref. \cite{Zu}. In comparing our scheme with these previous ones, they all use the three-level systems, and the one-qubit gates in our scheme can take the same scales of physical parameters as those in the previous schemes. The only difference is that arbitrary one-qubit gates in our scheme can be realized by using a single-shot implementation while the previous ones need to combine two sequentially implemented gates in general. The one-qubit gates in our scheme are naturally compatible with previously proposed two-qubit non-adiabatic holonomic gates, and therefore universal quantum computation can be realized. In other words, in order to realize universal quantum computation, the non-adiabatic holonomic two-qubits gates in our scheme can be constructed just by the same approaches as used in the previous works, while the one-qubits gates are constructed by the new approach.

In conclusion, we put forward a novelty scheme by which one can directly realize an arbitrary holonomic one-qubit gate with a single-shot implementation. Based on a three-level model driven by laser pulses, we show that any holonomic one-qubit gate can be realized by separately varying the detuning, amplitude, and phase of lasers. Comparing with previous schemes, which need at least two sequential implementations, our scheme simplifies the procedures of realizing holonomic one-qubit gates, avoiding the extra work of combining two gates into one. The scheme can be implemented experimentally in various systems, such as trapped ions, superconducting qubits, or NV-centers in diamond. Our scheme of realizing arbitrary non-adiabatic holonomoc one-qubit gates is compatible with the previous schemes of realizing non-adiabatic holonomic two-qubit gates, and therefore the arbitrary holonomic one-qubit gates combining with the previous two-qubit gate can realize universal non-adiabatic holonomic quantum computation.  We also investigate the effects of some realistic errors. Our result shows that all the average fidelities decrease with the increasing of error coupling parameters. The average fidelity for dephasing depends only on the axis parameter $\alpha$ but not on $\beta$, while the average fidelity for pulse area error or frequency detuning error is totally independent of the rotation axis parameters. The details of the errors¡¯ effects on the gates with various rotation angles are given too.

\section*{Acknowledgments}
This work was supported by the National Basic Research
Program of China through Grant No. 2015CB921004. G.F.X. acknowledges support from the Fundamental Research Funds of Shandong University
under Grant No. 11160075614015. C.L.L. and P.Z.Z. acknowledge support from NSF China through Grant No. 11175105.

\end{document}